\documentclass[twocolumn,showpacs,preprintnumbers,amsmath,amssymb]{revtex4-1}

\usepackage{float}
\usepackage{graphicx}
\usepackage{dcolumn}
\usepackage{bm}
\usepackage{color,soul}

\begin{document}


\title{Drop spreading and gelation of thermoresponsive polymers}

\author{R. de Ruiter$^1$, L. Royon$^2$, J.H. Snoeijer$^{1,3}$ and P. Brunet$^2$}
\email{philippe.brunet@univ-paris-diderot.fr}
\affiliation{$^1$ Physics of Fluids Group and J. M. Burgers Centre for Fluid Dynamics, University of Twente, P.O. Box 217, 7500 AE Enschede, The Netherlands \\
$^2$ Laboratoire Mati\`ere et Syst\`emes Complexes UMR CNRS 7057, 10 rue Alice Domon et L\'eonie Duquet 75205 Paris Cedex 13, France \\
$^3$ Department of Applied Physics, Eindhoven University of Technology, P.O. Box 513, 5600MB
Eindhoven, The Netherlands}
\date{\today}

\begin{abstract}

Spreading and solidification of liquid droplets are elementary processes of relevance for additive manufacturing. Here we investigate the effect of heat transfer on spreading of a thermoresponsive solution (Pluronic F127) that undergoes a sol-gel transition above a critical temperature $T_m$. By controlling the concentration of Pluronic F127 we systematically vary $T_m$, while also imposing a broad range of temperatures of the solid and the liquid. We subsequently monitor the spreading dynamics over several orders of magnitude in time and determine when solidification stops the spreading. It is found that the main parameter is the difference between the substrate temperature and $T_m$, pointing to a local mechanism for arrest near the contact line. Unexpectedly, the spreading is also found to stop below the gelation temparature, which we attribute to a local enhancement in polymer concentration due to evaporation near  the contact line.

\end{abstract}


\maketitle                              

\section{INTRODUCTION}

Many applications require the controlled deposition of minute amounts of liquid on solid substrates. The recent massive development of  3D-printing by continuous addition of matter has raised issues on the feasibility of various processes \cite{Lipson_Kurman13}. In one of them, liquid with complex rheology is extruded through a narrow slit and forced to spread on a solid \cite{Morissette00,Lewis06}, where solidification has to occur promptly and in a reproducible fashion, in order to avoid detrimental mechanical properties of the final product. Solidification itself can be induced by heat transfer at the substrate, simply because during extrusion the liquid temperature is above the melting temperature. This involves the spreading of a small amount of liquid which is susceptible to become solid over a comparable time-scale: such a situation involves coupling between dynamical wetting, heat transfer and rheology in a non-trivial way. Such configuration also finds applications in optical lens manufacturing \cite{Roy16}.

How does heat transfer influence the dynamics of drop spreading, and what determines the final droplet shape once solidification is completed? Arrest of the drop can be due to temperature-dependent viscosity (generally diverging when approaching the melting point), or more dramatically due to solidification near the liquid/solid interface. Such cases were investigated for droplet impact~\cite{Schiaffino_Sonin97_1,Schiaffino_Sonin97_2,Schiaffino_Sonin97_3,Jalaal_Stoeber14,Beesabathuni15}, as well as for spontaneous spreading after gentle deposition on solid substrates~\cite{Tavakoli14,Tavakoli15,deRuiter17}. Inkjet printing and applications such as spray coating indeed involve droplet impact. When the drop's kinetic energy is sufficiently large this can induce splashing~\cite{Thoroddsen12,ARFM_Josserand16} or undamped capillary waves \cite{Vadillo09} that can be detrimental for the final product. Therefore, understanding the solidification after smooth and gentle deposition at negligible inertia is of great importance for applications of additive material processing. Even without inertia, however, the problem is very intricate since the wedge-like geometry near the contact line leads to singularities of viscous dissipation~\cite{Huh_Scriven,Bonn09,Snoeijer13}, as well of heat and evaporative fluxes~\cite{Deegan97,Deegan_pre00,Berteloot_epl08,Monteux_epl09}.

In this study we experimentally investigate the dynamics of drop spreading of thermoresponsive polymer solutions that undergo a gelation transition. This is a model system for various applications involving cross-linking polymers, for which solidification and spreading occur within a comparable time scale. Hence, during spreading the liquid locally solidifies by gelation, which leads to a radius of arrest that is smaller (and to a contact angle that is larger) than in the absence of gelation. Our goal is to quantify the time after which solidification dominates over spreading, by determining the time and radius of arrest of the drop, under different conditions of substrate temperature $T_0$. The choice of liquid is crucial, as it has to enable an accurate and convenient control of $T_0$ close to $T_m$. In this sense, experiments with water are inconvenient as droplets condense on the substrate when $T_0$ approaches $T_m$. Furthermore, water undergoes sub-cooling, which adds complexity to the freezing process \cite{Tavakoli15}. Therefore, we opted for solutions of Pluronic F127, which is a thermoresponsive co-polymer that undergoes a cross-linking gelation transition above a given temperature (without latent heat). The same system was recently used to study drop impact \cite{Jalaal_Stoeber14} and the gelation process during spreading \cite{Jalaal_Seyfert18}. Here we focus on the detailed spreading dynamics, the parameters that control the gelation-induced contact line arrest and the role of relative humidity.

The paper is organised as follows. We first present the experimental setup and rheological properties of the liquid (II), then we show quantitative results on the spreading dynamics under various conditions (III). We accurately record the short-time and long-time dynamics of the spreading radius of the droplet after it contacts the solid substrate. We vary the substrate temperature over a broad range of temperatures, from far below $T_m$ to far above $T_m$. We carry out these experiments for different mass concentration of polymer in water, which changes the value of $T_m$. Then, we discuss these results in light of previous works and modeling (IV) and we conclude by a summary and outlook of our main findings (V).

\section{DESCRIPTION OF THE EXPERIMENTS}

\subsection{The drop spreading setup}

\begin{figure}
\includegraphics[scale=0.38]{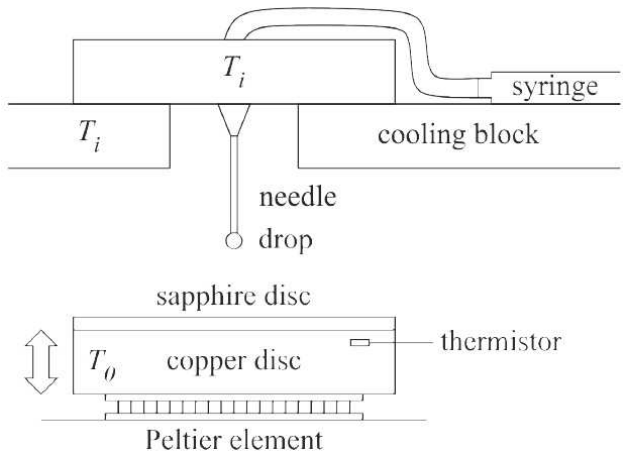}
\caption{Sketch of the experimental setup, see text for details. A millimeter-sized drop of temperature $T_i$ is deposited with a needle at $V$=0 onto a conductive substrate of temperature $T_0$. The drop then spreads until it reaches its final radius, while the dynamics is recorded with a high-speed camera.}
\label{fig:setup}
\end{figure}

The setup used for the temperature-controlled spreading experiments is sketched in Figure \ref{fig:setup}. By infusing a syringe with liquid, a liquid drop is slowly inflated at the tip of a needle of outer diameter $d_n=$ 650 $\mu$m, which leads to a drop radius $R_0= 860$ $\mu$m $\pm$60. Subsequently, the drops are gently brought in contact with a substrate of controlled temperature $T_0$. The substrate is a sapphire disc of thickness 2 mm. The choice for sapphire allows us to operate on a substrate that is both thermally conductive ($\kappa_s$ = 32 W/(m K)) and smooth. Its heat capacity $C_p$ = 0.76$\times$10$^{3}$ J / (kg K) and density $\rho_s$ = 3.98$\times$10$^{3}$ kg.m$^{-3}$ yield a thermal diffusivity $\alpha_s = \frac{\kappa_s}{C_p \rho_s}$ = 1.058$\times$10$^{-5}$ m$^{2}$s$^{-1}$. The sapphire disc is put on top of a copper disc, which temperature is actively regulated with a Peltier element and thermistor. In this way we control the (bottom) substrate temperature $T_0$ from 2$^{\circ}$C to 50$^{\circ}$C with a 0.1$^{\circ}$C accuracy. To ensure a well controlled initial temperature $T_i$ for the drop, the syringe is attached to a cooling block and insulated from the atmosphere with insulating foam. The spreading dynamics is acquired with a high-speed camera (Photron SA-X2) on which a telecentric zoom lens (Navitar 12X) is mounted. A typical resolution of 6.2 $\mu$m/pixel ensures the precise extraction of the whole dynamics.

Unless stated otherwise, the ambient temperature and the relative humidity are those of the room ($T_a$ = 20$^{\circ}$C, RH measured between 50\% and 60\%). Some experiments were carried out at lower or higher RH, by using a glass cuvette to cover the droplet area. Inside the cuvette, dry nitrogen (N$_2$) or water vapor-enriched air was flushed to fix the RH respectively to around 10\% and 90\% - 100\%. 

In order to minimize the heat exchange between the drop and the atmosphere during the approaching phase, especially in situations where $T_a \ge T_m$ (high polymer concentration), experiments were performed according to the following protocol. First, we flushed the needle with several drops, and let the last one hang on the needle, in order to have the drop temperature as close as possible to that in the syringe. Then, we gently, but quickly enough, raised the sapphire substrate. The entire process never lasted more than 3 seconds. Finally, due to the relatively high conductivity of the sapphire substrate, the solid temperature can be assumed to remain constant during spreading.

\begin{table}[htp]
\caption{Gelation temperature $T_m$ of the Pluronic solutions at the different mass percentages used in experiments. The different experimental conditions are indicated with cross 'x' symbols, where $\Delta T_i = T_i - T_m$, where $T_i$ is the temperature at which liquid is injected.}
\begin{center}
\begin{tabular}{c|c|clclclclclc}
\hline
Conc.  & Gelation temp. &  &   &   &    &   \\
\hline
\hline

$c_0$ (\%)  & $T_m$ ($^{\circ}$C) &\, $\Delta T_i$ = &\, -1  &\,  -5  &\, -10  & \, -15  \\
\hline
16.7 & 28.9  &  & - & - &\, x &\, - \\
20 & 22.0 &  & x & x &\, x &\, x \\
25 & 16.5 &  & - & x &\, x &\, - \\
30 & 11.5 &  & - & x &\, - &\, -  \\
\hline
\end{tabular}
\end{center}
\label{tab:properties}
\end{table}%

\subsection{Rheological properties of the liquid}

\begin{figure}
\includegraphics[scale=0.32]{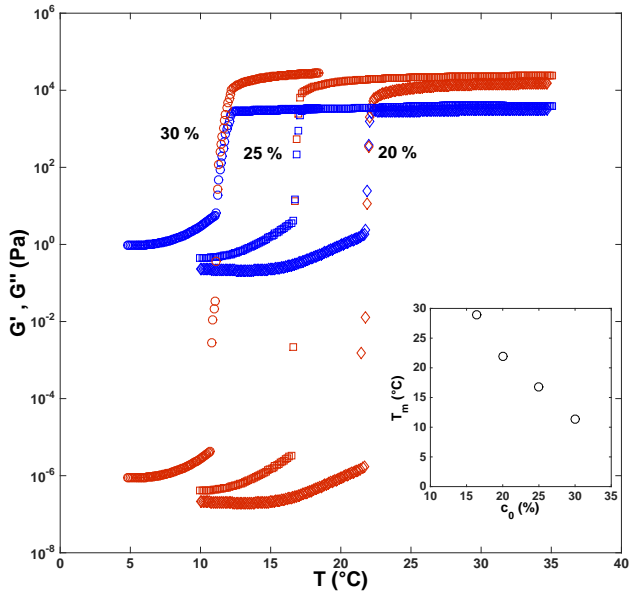}
\caption{Rheological properties of the liquids. Storage modulus $G'$ (red markers) and loss modulus $G''$ (blue markers) versus temperature for 3 different co-polymer concentrations, $c_0$ = 20 \% (diamonds), 25\% (squares) and 30 \% (circles). Insert: Gelation temperature $T_m$ versus co-polymer concentration. Experiments were carried out at a shear-rate of 10 s$^{-1}$, corresponding to a frequency $\omega$ of 10 Hz.}
\label{fig:rheo1}
\end{figure}

The experiments were performed with Pluronic F127 from BASF\copyright (PEO$_{106}$PPO$_{70}$PEO$_{106}$), which is an aqueous solution of thermoresponsive co-polymer. This system, also considered in \cite{Jalaal_Stoeber14}, allows one to operate at ambient atmosphere and temperature, enabling precise visualization and comfortable experimental conditions. This co-polymer undergoes a transition to a gel phase due to cross-linking of micelles formed by the molecules, when the temperature gets higher than a threshold $T_m$ that depends on the mass percentage $c_0$ of polymer in water. This is a second-order phase transition and therefore there is no latent heat associated to the sol-gel phase transition, which simplifies the solidification process.

\begin{figure*}
\includegraphics[scale=0.3]{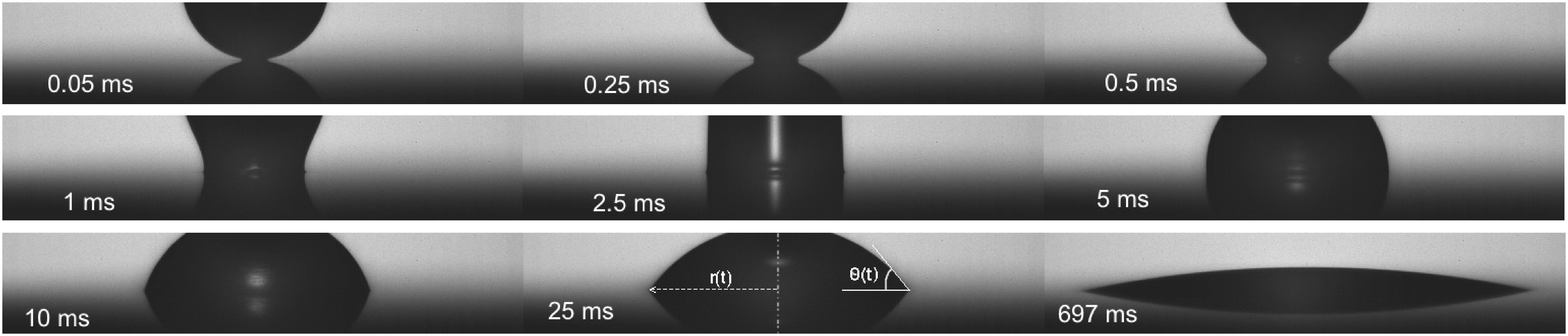}
\caption{Sequence of successive snapshots showing the drop spreading on a substrate with temperature $T_0$= 10$^{\circ}$C well below the gelation temperature $T_m$=22$^{\circ}$C ($c_0$ = 20 \%). This sequence shows a late solidification-induced pinning and large arrest radius (spreading has stopped in the last image). Time after contact is labelled on the snapshots. The basal radius $r (t)$ and contact angle $\theta (t)$ can be extracted from these images.}
\label{fig:drop_spreading_sequence1}
\end{figure*}

\begin{figure*}
\includegraphics[scale=0.3]{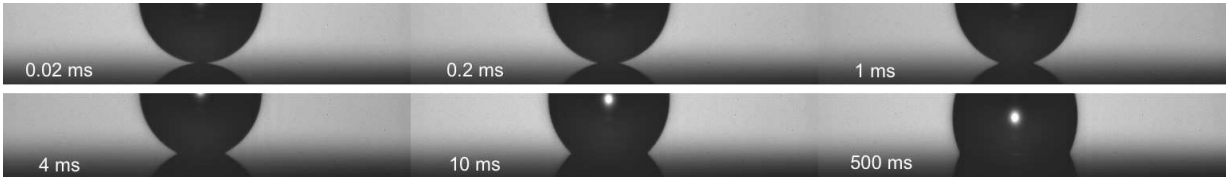}
\caption{Similar sequence as in Fig.~\ref{fig:drop_spreading_sequence1} but for a substrate temperature $T_0$=30 $^{\circ}$C, well above the gelation temperature $T_m$=22$^{\circ}$C. This sequence shows early pinning and small arrest radius. Time after contact is labelled on the snapshots.}
\label{fig:drop_spreading_sequence2}
\end{figure*}

The system represents a relatively complex rheology, though it has been thoroughly characterised in various studies \cite{Prudhomme96,Habas04,Chaibundit07,Boucenna10}. Here we independently quantify the viscoelastic properties of Pluronic F127 solutions. We employ a dynamic rheology technique, using a Physica MCR 500 rheometer (Anton Paar) with a cone and plate geometry (50 mm diameter with a 1$^{\circ}$ cone angle). Prior to the measurements, a strain sweep test was conducted within the linear viscoelastic strain range in the frequency range of 0.01-100 Hz. The rheological behavior of each sample was measured by performing a temperature sweep. The temperature dependences of storage modulus $G'$ and loss modulus $G''$ were measured by heating the sample from 5$^{\circ}$C to 35$^{\circ}$C. The rate was 1$^{\circ}$C/min during temperature scans, with no notable difference between heating and cooling. The deformation was fixed at 1\%, which was in good agreement with the linear viscoelastic region for all the samples.

The results are shown in Figure \ref{fig:rheo1} for three concentrations $c_0$ = 20\%, 25\% and 30\%, representing the mass percentage of Pluronic F127 in water. The presented data correspond to a frequency $\omega=10 $Hz. We observe a remarkably sharp increase of both the elastic modulus ($G'$, red markers) and of the loss modulus ($G''$, blue markers) at a well-defined temperature $T_m$. This increase is by more than 10 orders of magnitude for $G'$ and by almost 4 orders for $G''$. Interestingly, above $T_m$ the value of $G'$ is at least one order of magnitude higher than for $G''$ such that the elastic behaviour will dominate over viscous dissipation. These observations allow us to very sharply identify the sol-gel transition temperature. The inset in Fig. \ref{fig:rheo1} shows the dependence of $T_m$ on the concentration $c_0$, with values similar to previous measurements \cite{Prudhomme96,Habas04,Chaibundit07,Boucenna10,Jalaal_Cottrell2017}. The dynamic viscosity can be deduced as $\eta = \frac{G''}{\omega}$. Here, it is important to note that below the transition, the viscosity of the liquid only slightly increases with temperature, but remains below 10 cP. Therefore, we expect that below the gelation point, the liquid will spread in the inertia-capillary regime at early times, as previously reported for low-viscosity fluids \cite{Biance04,Bird08,Amberg12,Winkels12,Eddi13}.

\section{RESULTS}

\subsection{Dynamics}

Three temperatures could play a role for the spreading dynamics and final drop shape: the melting (or gelation) temperature $T_m$, which can be tuned by $c_0$ (see Table 1 and Fig.~\ref{fig:rheo1}), the substrate temperature $T_0$ and the initial liquid temperature $T_i$. Operating with different $T_i$ allows us to investigate the possible lag induced by the heat transfer required to raise the liquid temperature during spreading. Therefore, we define two distinct temperature differences as relevant control parameters

\begin{equation}
\Delta T_0 = T_0 - T_m, \quad \quad \Delta T_i = T_i - T_m.
\end{equation}
These respectively are the difference between substrate temperature and gelation temperature, and the difference between the initial liquid temperature and the gelation temperature. Typical movies for $\Delta T_0 < 0$ and $\Delta T_0 >0$ are presented in Figs.~\ref{fig:drop_spreading_sequence1} and \ref{fig:drop_spreading_sequence2}. When the substrate temperature is below the solidification threshold, the drop nearly completely spreads out over the substrate ($\Delta T_0 < 0$, Fig.~\ref{fig:drop_spreading_sequence1}). However, we observe that the spreading \emph{does} stop at a finite radius. This is surprising since Pluronic F127 in solution acts as a surfactant and one would have expected a complete wetting on sapphire substrates. When the substrate temperature is above the solidification threshold ($\Delta T_0 > 0$, Fig.~\ref{fig:drop_spreading_sequence2}), the solidification occurs in the early stage of spreading and the final result is a droplet of small basal radius and large contact angle. 

\begin{figure*}[!]
\includegraphics[scale=0.55]{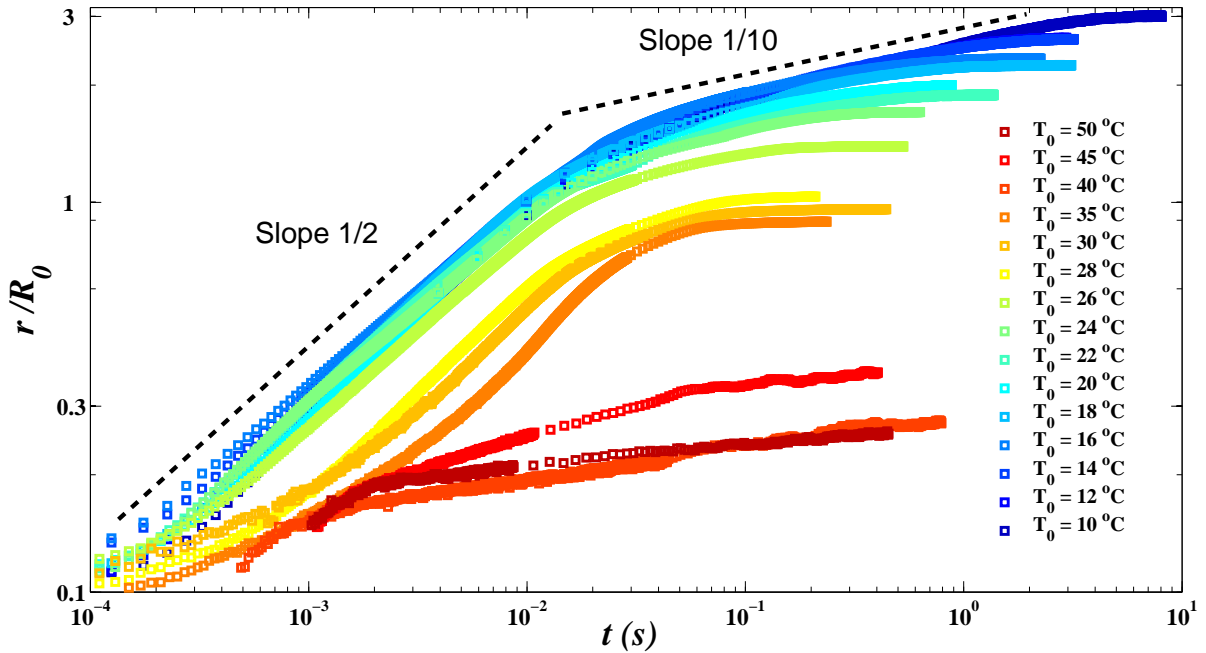}
\caption{Spreading radius $r$ (normalised by the initial drop radius $R_0$) as a function of time after contact $t$, for various substrate temperatures $T_0$. The concentration was kept fixed at $c_0 =$ 20\% ($T_m = 22 ^{\circ}C$) and each experiment had the same initial liquid temperature $T_i = 7 ^{\circ}C$ ($\Delta T_i = - 15 ^{\circ}C$). Colors code for $T_0$, from cold (blue) to the warm (red). Dotted lines indicate two power laws (1/2 and 1/10) for $T_0 \le T_m$ whereas the spreading dynamics does not follow any power-law for $T_0 \ge T_m$.}
\label{fig:radius_vs_time1}
\end{figure*}

\begin{figure}
\includegraphics[scale=0.35]{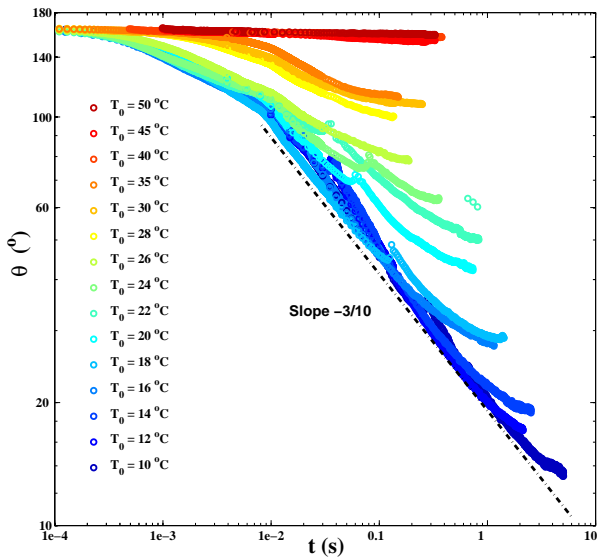}
\caption{Contact angle $\theta$ versus time, for various $T_0$ (same conditions as in Fig.~\ref{fig:radius_vs_time1}).}
\label{fig:angle_vs_t}
\end{figure} 

\begin{figure}[t]
\includegraphics[scale=0.43]{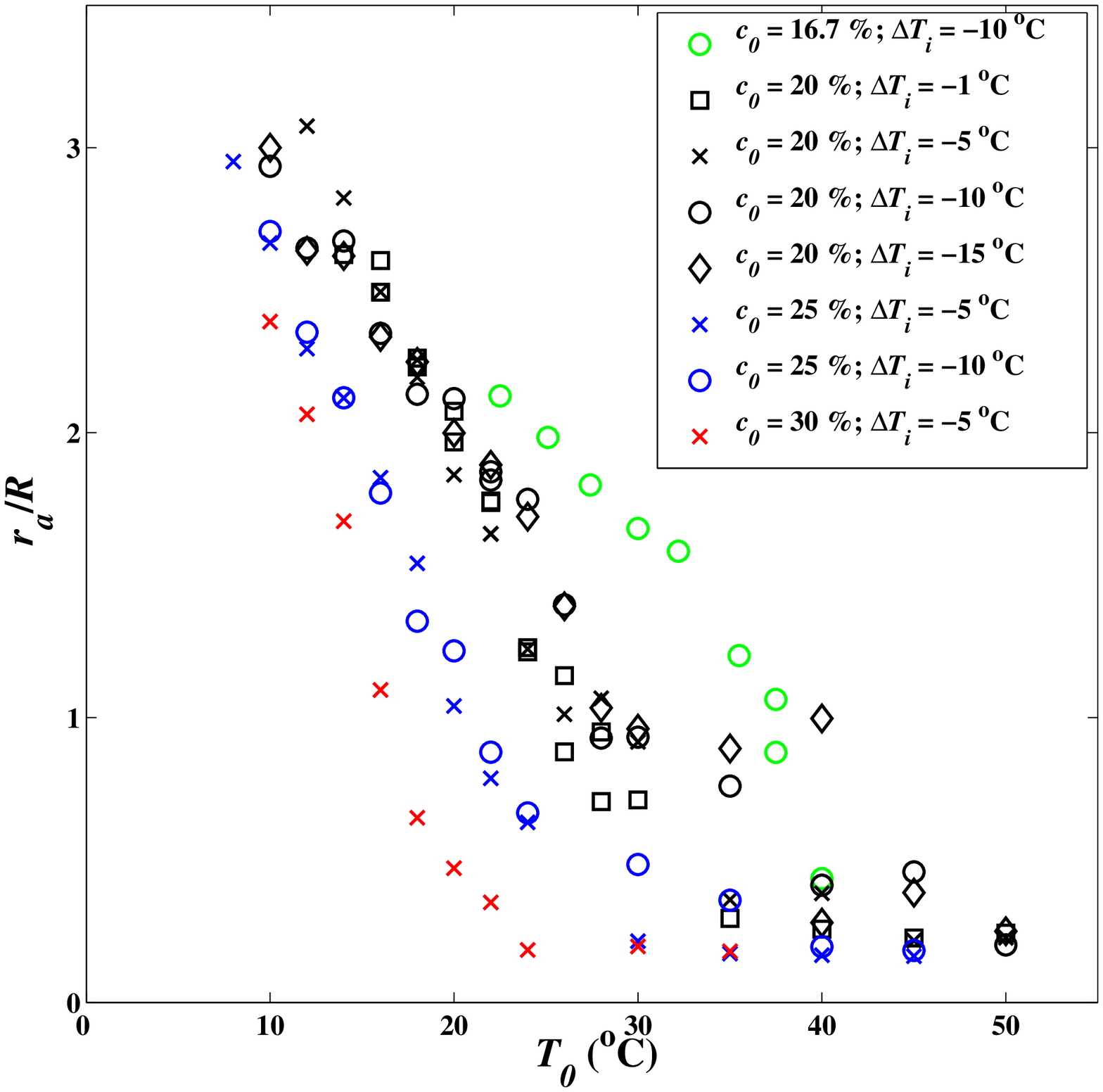}
\includegraphics[scale=0.43]{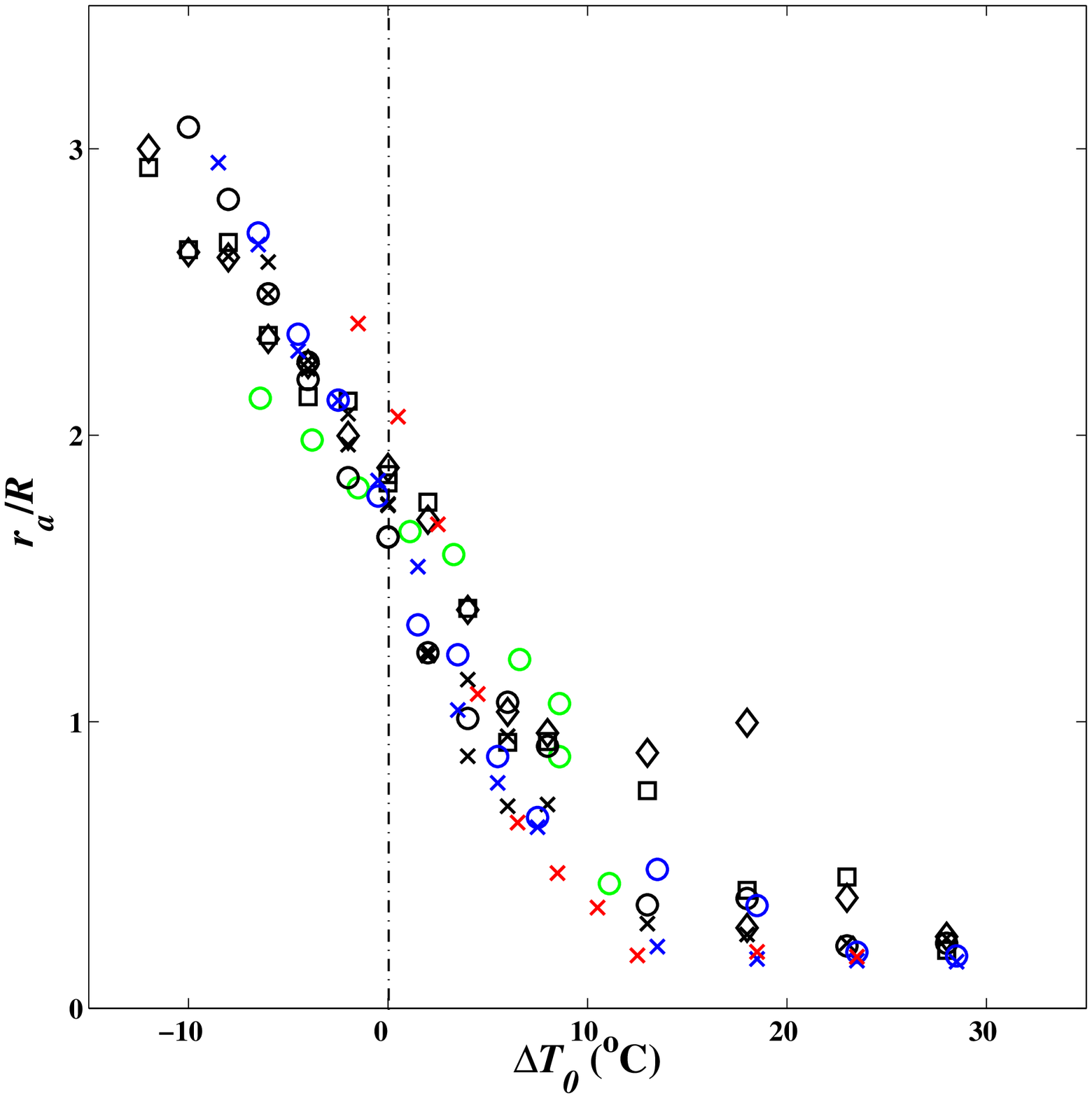}
\caption{\textit{Top} - Dimensionless (final) arrest radius versus substrate temperature $T_0$ for different gelation temperature $T_m$ (controlled via the Pluronic F127 concentration $c_0$) and different initial liquid temperature $T_i$. \textit{Bottom} Same data plotted versus $\Delta T_0 = T_0 - T_m$.}
\label{fig:radius_vs_T}
\end{figure}

To further quantify the spreading dynamics, and subsequent arrest by solidification, we extract the basal radius $r$ and contact-angle $\theta (t)$ (defined in Fig.~\ref{fig:drop_spreading_sequence1}) as a function of time after contact $t$. Figure \ref{fig:radius_vs_time1} shows the resulting $r(t)$ for various substrate temperatures $T_0$, but for the same concentration $c_0 = 20$ \% ($T_m = 22 ^{\circ}C$) and the same initial temperature $T_i = 7 ^{\circ}C$ ($\Delta T_i = - 15 ^{\circ}C$). The colors represent the temperature $T_0$, from blue (low $T_0$) to red (high $T_0$). The data in Fig. \ref{fig:radius_vs_time1} clearly show a change in spreading dynamics upon crossing the melting temperature. First, we see that low enough $\Delta T_0$ (in this specific case $\Delta T_0 < - 4^\circ$C), the spreading is consistent with $r \sim t^{1/2}$ at early times and $r \sim t^{1/10}$ in the final phases of spreading. These are the usual spreading characteristics for isothermal liquid drops, for which the exponent 1/2 corresponds to inertial spreading \cite{Biance04,Bird08,Amberg12,Winkels12,Eddi13}) while 1/10 is the classical Tanner's law for viscous spreading \cite{Tanner79,Hocking94}. The dynamics above the solidification threshold ($\Delta T_0 >0$) is very different and does not exhibit these exponents. Instead, a smooth and slow increase of $r$ is followed by a phase where the radius reaches its final contant value. This shows that the local freezing near the contact line induces pinning on the substrate. The trends observed in Fig.~\ref{fig:radius_vs_time1} are found for all tested concentrations $c_0$ and for all initial temperatures $T_i$ (see Table 1). 

The same trends can be observed when monitoring the evolution of the contact angle $\theta (t)$ during the spreading experiment. Figure \ref{fig:angle_vs_t} shows $\theta(t)$ under the same conditions as in Fig.~\ref{fig:radius_vs_time1}. It confirms that increasing the substrate temperature slows down the dynamics, and finally leads to ``taller" drops of higher contact angles. Once again, for substrate temperatures well below the threshold of solidification we recover the usual drop spreading characteristic, here represented by a decrease of contact angle with a -3/10 exponent in Tanner's regime (see also \cite{deRuiter17}).

\subsection{Condition of arrest}

We now quantify the conditions of arrest by determining the final spreading radius $r_{\text{max}}$ for different experimental conditions. In particular we explore the dimensionless spreading parameter $r_{\text{max}}/R$, for various temperatures $T_0$  and $T_i$, and for various concentrations $c_0$. The results are shown in Fig. \ref{fig:radius_vs_T}. The main trends that can be extracted are that the arrest radius decreases upon increasing $T_0$ or $c_0$. This is to be expected, since both a high temperature and high concentration promote the solidification process. The curves in the top panel of Fig. \ref{fig:radius_vs_T} appear to cluster by color, i.e. by their concentration $c_0$. This suggests that the initial temperature, which was varied from $\Delta T_i= -1^{\circ}$C to -15$^{\circ}$C, is not an important parameter for the solidification process. This important observation is further confirmed in the lower panel of Fig. \ref{fig:radius_vs_T} where we report the same data but now as a function of $\Delta T_0=T_0 - T_m$. This accounts for the differences in the solidification temperature $T_m$ for the various concentrations $c_0$. The data exhibit a good collapse and thus reveal that $\Delta T_0$ is indeed the key control parameter. One can therefore conclude that the time lag necessary to pre-heat the liquid from $T_i$ to $T_m$ is negligible compared to the time of spreading before pinning occurs. Only for the highest $T_0$, for which the time of spreading is considerably shorter, the time lag for pre-heating has some effect; this can be inferred from the scatter at high temperatures in the bottom panel of  Fig. \ref{fig:radius_vs_T}. 

\begin{figure}[!b]
\includegraphics[scale=0.43]{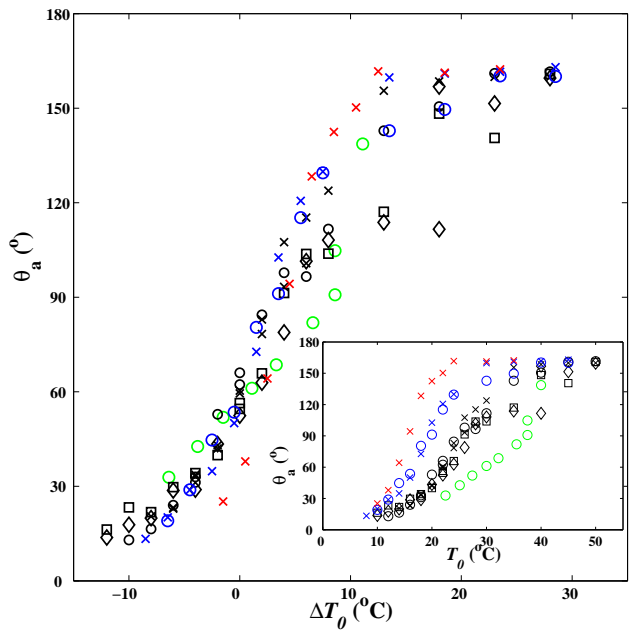}
\caption{Arrest contact angle versus temperature difference $\Delta T_0 = T_0 - T_m$. Inset: arrest contact angle versus $T_0$. Symbols are similar to Fig.~\ref{fig:radius_vs_T}.}
\label{fig:angle_arrest_versus_DeltaT}
\end{figure}

We also report the data in terms of contact angles in Fig. \ref{fig:angle_arrest_versus_DeltaT}, where we show the arrest contact angle $\theta_{\text{arrest}}$ versus $\Delta T_0$. It is found that the contact angle can be controlled over a remarkably large range, from about 15$^{\circ}$ to 160$^{\circ}$. This will be of interest in applications where a control of the droplet shape is required. The sharpest increase of the contact angle is measured around $\Delta T_0$=0, while a saturation appears at $\Delta T_0\ge 15^{\circ}$, when the substrate temperature is warm enough to induce prompt solidification. The inset in Fig.~\ref{fig:angle_arrest_versus_DeltaT} shows the same data but plotted versus $T_0$. This again confirms that $\Delta T_0$ is the relevant control parameter, while $T_i$ has a subdominant effect, only at larger temperature. 

\section{INTERPRETATION}

The solidification of the entire drop takes several seconds. This experimental observation can be understood from heat conduction, which gives a timescale $\tau \sim R^2/\alpha$, where $\alpha$ = 0.147$\times$10$^{-6}$ m$^2$.s$^{-1}$ is the  thermal diffusivity of water. Indeed, for a millimeter-sized droplet this gives a solidification time $\tau \simeq$ 7 s. Since most of the experiments exhibit arrest times well below 1 second, the arrest must be induced at scales much smaller than the droplet size. This is corroborated by the observation that the initial drop temperature $T_i$ is irrelevant, suggesting that the liquid close to the substrate very quickly adapts to the temperature of the substrate. Indeed, owing to the large difference between substrate and liquid thermal conductivities ($\kappa_s \gg \kappa_l$), one would expect the temperature at the contact surface (and in a thin enough layer just above the contact surface), to equal that of the substrate.

Still, the results in Figs. \ref{fig:radius_vs_T} and \ref{fig:angle_arrest_versus_DeltaT} are remarkable in several regards. In contrast to the dramatic change in rheology around $T_m$, the final radius and final contact angle exhibit a very smooth dependence on substrate temperature, even when $T_0 \approx T_m$. From this perspective, there are two observations that need to be clarified. For temperatures $T_0 > T_m$ one might naively expect a nearly instantaneous solidification at the point of contact, in which case there would be hardly any contact line motion at all. Yet, drops spread substantially for substrates even well above the gelation temperature. A second open issue is why for temperatures $T_0 < T_m$ the contact line pins, as opposed to the expected complete spreading of the droplet. Below we comment on both aspects and we provide experimental evidence that evaporation plays a role in the arrest of spreading.

\subsection{Mechanism of solidification-induced pinning}

We first discuss the question of what causes the delay in spreading arrest for substrate temperatures above $T_m$. As already mentioned in \cite{Jalaal_Stoeber14}, the problem at hand is closely related to the spreading of common liquids on a substrate colder than the freezing temperature. The gelation occurring above $T_m$ is analogous to liquid freezing below $T_m$, albeit without the release of latent heat. 

For freezing-induced pinning, the literature offers various hypotheses for the criterion for contact line arrest~\cite{Schiaffino_Sonin97_1,Schiaffino_Sonin97_2,Schiaffino_Sonin97_3,Tavakoli14,Tavakoli15,deRuiter17}. For example, the delay in the arrest was argued to be caused by the requirement for a critical nucleus to form near the contact line \cite{Tavakoli14,Tavakoli15}. However, a detailed description of growth of the nucleus in the presence of flow is lacking. Alternatively, for the spreading of hexadecane drops this freezing-delay was attributed to the effect of kinetic undercooling~ \cite{deRuiter17}. Kinetic undercooling leads to a shift in freezing temperature if the liquid velocity of the freezing front is nonzero \cite{Davis_Solidification}. The same mechanism could be at play here, since the contact line velocity is very high at the start of the experiment, preventing immediate solidification. The condition for spreading arrest would be that the velocity of the (freezing or gelation) front exceeds the velocity of the contact line. To confirm, or rule out, the effect of kinetic undercooling in spreading arrest, there is a need for an independent measurement of the so-called kinetic coefficient that relates the freezing temperature and front velocity. 

For both mechanisms discussed above the solidification is assumed to be initiated at the contact line. This is recently confirmed explicitly for slowly spreading Pluronic F127 \cite{Jalaal_Seyfert18}. In these experiments the temporal evolution of the gelation front inside the drop was resolved, displaying several interesting features such as the formation of a thin crust over the free surface. Indeed, the very first gel was observed to form at the contact line, albeit only after the contact line was pinned.

\subsection{Influence of relative humidity and  evaporation}

\begin{figure}
\includegraphics[scale=0.5]{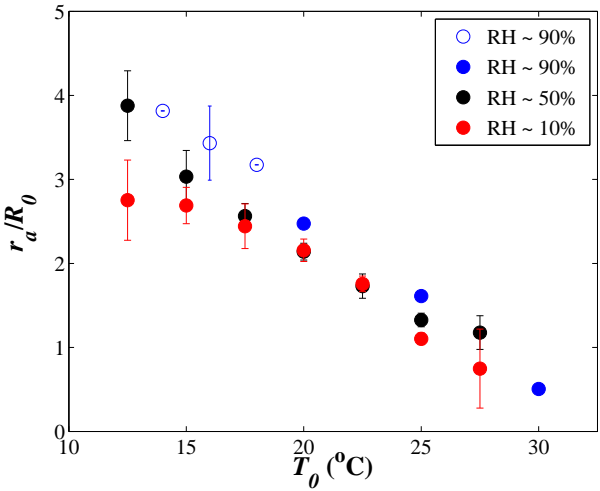}
\caption{Dimensionless radius of arrest versus $T_0$ for three different \%RH : black symbols represent ambient conditions, i.e. RH between 50 and 60\% and red symbols represent low humidity conditions (RH = 10 \%), while blue symbols represent very humid conditions close to saturation (RH $ = $ 90 \% $\pm$ 5). Open symbols represent situations when spreading was still slowly going on, leading to an underestimation of $r_a/R_0$. Other experimental conditions are identical for all data points: $c_0$ = 20 \% ($T_m = 22^{\circ}$C) and $\Delta T_i = - 5 ^{\circ}$.}
\label{fig:influ_rh}
\end{figure}

\subsubsection{Experiment}

We now turn to the observation that the drop spreading arrests even when $T_0 \le T_m$. Here we show that this unexpected effect is due to water evaporation near the contact line. Though the timescale of evaporation of the entire drop is much larger than the typical spreading time, we recall that the evaporation flux is strongly enhanced near the contact line. The rate of evaporation even diverges for contact angles below 90$^\circ$~\cite{Deegan97,Deegan_pre00,Berteloot_epl08,Monteux_epl09}. The role of solvent evaporation for gel formation in Pluronic was also invoked in \cite{Jalaal_Seyfert18}. For all these reasons we experimentally verified the possible influence of \%RH on the spreading dynamics and final radius.

We reproduced a series of experiments, previously carried out under ambient humidity (measured between 50 and 60 \% during all the experiments), but now while covering the surrounding of the substrate with a transparent cuvette. By flushing with a gentle stream of dry nitrogen, we could operate at RH as low as 10 \%. To operate at high humidity condition, the cuvette was flushed with humidified air, with the help of a home-made humidifier. The RH inside the cuvette was continuously measured and the spreading experiment started as soon as we reached the maximal RH (ranging between 90 and 100 \%) before condensation droplets on the substrate were observed.

Figure \ref{fig:influ_rh} shows the arrest radius under the three conditions (dry, ambient, and humid) for different substrate temperatures $T_0$. It turns out that the drop stops spreading at smaller radius under dry conditions compared to ambient conditions. Furthermore, at high RH close to saturation the arrest radius is significantly higher than for ambient conditions. However, this difference is only significant for $T_0 \le T_m$. Around or far above $T_m$, no significant difference could be noticed between the three RH conditions. 

\subsubsection{Estimations based on scaling}

\begin{figure}
\includegraphics[scale=0.65]{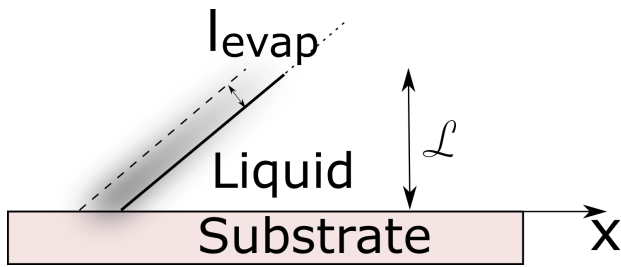}
\caption{Sketch of liquid evaporation near the wedge.}
\label{fig:sketch_evap}
\end{figure}

We now rationalise that the local evaporation, and the subsequent increase of polymer concentration near the contact line, is indeed a plausible mechanism for the observed arrest below $T_m$.  For this we consider a small volume of liquid near the contact line of characteristic size $\mathcal L$ as sketched in Fig.~\ref{fig:sketch_evap}. We estimate the water mass that is evaporated in this region during a typical experimental timescale $\tau \sim 1$~s. This loss of water can be computed from the diffusion-induced flux per unit width of contact line $J(x)$, where $x$ denotes the distance from the contact line (cf. Fig.~\ref{fig:sketch_evap}). From Deegan and colleagues' pioneering work on evaporation of colloidal droplets \cite{Deegan97,Deegan_pre00}, and from various other studies \cite{Berteloot_epl08,Monteux_epl09,Gelderblom_pre11}, $J(x)$ can be computed as 

\begin{equation}
J(x) \sim \frac{D (c_{\infty} - c_s)}{R^{1-\alpha} x^{\alpha} \rho_l} = U_0 \left( \frac{R}{x}\right)^\alpha,
\label{eq:j}
\end{equation}
where $D$ is the mass diffusion coefficient of water in air, $\rho_l$ the liquid mass density, and $c_{\infty}$ and $c_s$ respectively stand for the mass concentration of water vapor at infinity and near the drop, while $R$ is the drop-size. These quantities are collected in a characteristic velocity $U_0 = D\Delta c/R\rho_l$, which in our experiment is approximately $U_0  \sim 2.5\times10^{-7}$ m.s$^{-1}$~\cite{estimation}. 
The exponent $\alpha$ is known to depend on the contact angle: for $\theta=\pi/2$ the evaporative flux is uniform ($\alpha$ = 0), while it diverges close to the contact line for smaller angles ($\alpha \simeq 1/2$ for $\theta \ll 1$). 

As a first step, we simply assume the evaporative flux to be uniform, $J \sim U_0$ and estimate the relative change of volume in a time $\tau$ as

\begin{equation}
\mathcal{F} = \frac{l_{\text{evap}}}{\mathcal{L}} \sim \frac{U_0 \tau}{\mathcal{L}}.
\label{eq:f}
\end{equation}
Here $l_{\text{evap}}$ is the typical length evaporated off in a time $\tau$ (Fig.~\ref{fig:sketch_evap}). Now, we can ask what the size $\mathcal L$ is, if we assume that the concentration increase necessary to induce gelation is about 10\%. This corresponds to $\mathcal F \sim 0.1$, which gives a typical size $\mathcal L \sim 2.5\, \mu\textrm{m}$. Following the same argument but now taking into account that the flux is non-uniform, with $\alpha=1/2$ in (\ref{eq:j}), we obtain

\begin{equation}
\mathcal{F}  \sim \frac{U_0 \tau}{\mathcal{L}} \left( \frac{R}{\mathcal L}\right)^{1/2},
\label{eq:f2}
\end{equation}
where we integrated the flux from the contact line to a length $\mathcal L$. If we now again estimate the size of the corner based on $\mathcal F \sim 0.1$, we obtain $\mathcal L \sim 18\, \mu\textrm{m}$. These arguments show that, indeed, the effect of evaporation at a small distance from the contact line could lead to a sufficient enhancement of the concentration -- ultimately leading to arrest even when $T_0 < T_m$.

\vspace{0.5 cm}

\section{Conclusion}

In conclusion, we conducted experiments of spontaneous spreading of a drop of thermo-responsive liquid Pluronic F127, experiencing gelation via cross-linking of micelles above $T_m$, on substrates of various temperature $T_0$ from far below to far above $T_m$. We varied the polymer concentration, which allowed to vary $T_m$. For substrate temperatures well below the gelation temperature ($T_0 \le T_m - 4$), we recover the crossover between an early inertio-capillary stage and a late viscous-capillary stage. For higher $T_0$, no power-law could be identified in the spreading dynamics, which is due to the effect of heat transfer. Above a critical temperature (e.g. 10$^{\circ}$C for $c_0$ = 20 \%), the spreading stops to a final arrest radius which is the consequence of solidification-induced pinning at the contact line. 

The main results are: (1) the final radius is not dependent on initial liquid temperature, but only on the difference $T_0-T_m$, and (2) the final radius decreases with $T_0$, even in the range $T_0 \le T_m$. This last unexpected behavior can be attributed to evaporation near the contact line which leads to a local increase of the concentration. 
Both the final basal radius and contact angle can be controlled in a large range, which offers potential interest in lens manufacturing and additive 3D printing.

\acknowledgments{We thank Maxime Costalonga for his help on experiments at high humidity, and Maziyar Jalaal for discussions. P.B.'s stays in Twente's University were supported by the CNRS (PICS Program).}


\begin{thebibliography}{100} 

\bibitem{Lipson_Kurman13}
H. Lipson and M. Kurmann,
\textit{Fabricated: The New World of 3D Printing (John Wiley \& Sons, Indianapolis).}

\bibitem{Morissette00}
S.L. Morissette, J.A. Lewis, J.I. Cesarano, D.B. Dimos and T. Baer,
\textit{Solid freeform fabrication of aqueous aluminaÄ poly(vinyl alcohol) gelcasting suspensions,}
J. Am. Ceram. Soc. \textbf{83}, 2409-2416 (2000).

\bibitem{Lewis06}
J.A. Lewis,
\textit{Direct Ink Writing of 3D Functional Materials,}
Adv. Funct. Mater. \textbf{16}, 2193‚Äö√Ñ√¨2204 (2006).

\bibitem{Roy16}
A. C. Roy, M. Yadav, E. P. Arul, A. Khanna and A. Ghatak,
\textit{Generation of Aspherical Optical Lenses via Arrested Spreading and Pinching of a Cross-Linkable Liquid},
Langmuir DOI: 10.1021/acs.langmuir.5b04631 (2016).

\bibitem{Schiaffino_Sonin97_1}
S. Schiaffino and A. A. Sonin,
\textit{Motion and arrest of a molten contact line on a cold surface: An experimental study}, 
Phys. Fluids \textbf{9}, 2217?2226 (1997).

\bibitem{Schiaffino_Sonin97_2}
S. Schiaffino and A. A. Sonin,
\textit{On the theory for the arrest of an advancing molten contact line on a cold solid of the same material},
Phys. Fluids \textbf{9}, 2227 (1997).

\bibitem{Schiaffino_Sonin97_3}
S. Schiaffino and A. A. Sonin,
\textit{Molten droplet deposition and solidification at low Weber numbers},
Phys. Fluids \textbf{9}, 3172 (1997).

\bibitem{Jalaal_Stoeber14}
M. Jalaal and B. Stoeber,
\textit{Controlled spreading of thermo-responsive droplets,}
Soft Matt. \textbf{10}, 808 (2014).

\bibitem{Jalaal_Seyfert18}
M. Jalaal, C. Seyfert, B. Stoeber and N.J. Balmforth,
\textit{Gel-controlled droplet spreading,}
J. Fluid Mech. \textbf{837}, 115 (2018).

\bibitem{Beesabathuni15}
S. N. Beesabathuni, S. E. Lindberg, M. Caggioni, C. Wesner and A. Q. Shen,
\textit{Getting in shape: Molten wax drop deformation and solidification at an immiscible liquid interface},
J. Coll. Interf. Sci.  \textbf{445}, 231-242 (2015).

\bibitem{Tavakoli14}
F. Tavakoli, S. H. Davis, H. Pirouz Kavehpour,
\textit{Spreading and Arrest of a Molten Liquid on Cold Substrates},
Langmuir \textbf{30}, 10151 (2014).

\bibitem{Tavakoli15}
F. Tavakoli, S. H. Davis, H. Pirouz Kavehpour,
\textit{Freezing of supercooled water drops on cold solid substrates: initiation and mechanism}
J. Coat. Technol. Res. \textbf{12} 869-875 (2015).

\bibitem{deRuiter17}
R. de Ruiter, P. Colinet, P. Brunet, J. H. Snoeijer and H. Gerdelblom,
\textit{Contact line arrest in solidifying spreading drops},
Phys. Rev. Fluids \textbf{2}, 043602 (2017).

\bibitem{Tadrist}
S. David, K. Sefiane, L. Tadrist,
\textit{Experimental investigation of the effect of thermal properties of the substrate in the wetting and evaporation of sessile drops},
Coll. Surf. A: Physicochem. Eng. Aspects \textbf{298} 108-114 (2007).

\bibitem{Thoroddsen12}
S.T. Thoroddsen, K. Takehara and T.G. Etoh,
\textit{Micro-splashing by drop impacts,}
J. Fluid Mech. \textbf{706}, 560-570 (2012).

\bibitem{Vadillo09}
D. C. Vadillo, A. Soucemarianadin, C. Delattre and D. C. D. Roux
\textit{Dynamic contact angle effects onto the maximum drop impact spreading on solid surfaces,}
Phys. Fluids \textbf{21}, 122002 (2009).

\bibitem{ARFM_Josserand16}
C. Josserand and S. T. Thoroddsen,
\textit{Drop Impact on a Solid Surface},
Ann. Rev. Fluid Mech. \textbf{48}, 365-391 (2016). 

\bibitem{Huh_Scriven}
C. Huh and L. E. Scriven,
\textit{Hydrodynamic Model of Steady Movement  of a Solid / Liquid / Fluid Contact Line},
J. Coll. Interf. Sci. \textbf{71}, 85 (1971).

\bibitem{Bonn09}
D. Bonn, J. Eggers, J. Indekeu, J. Meunier, and E. Rolley, 
\textit{Wetting and spreading,} 
Rev. Mod. Phys. \textbf{81}, 739-805 (2009).

\bibitem{Snoeijer13}
J. H. Snoeijer and B. Andreotti, 
\textit{Moving contact lines: Scales, regimes and dynamical transitions,} 
Ann. Rev. Fluid Mech. \textbf{45}, 269-292 (2013).

\bibitem{Deegan97}
R. D. Deegan, O. Bakajin, T. F. Dupont, G. Huber, S. R. Nagel and T. A. Witten, 
Nature \textbf{389}, 827 (1997).

\bibitem{Deegan_pre00}
R. D. Deegan, O. Bakajin, T. F. Dupont, G. Huber, S. R. Nagel, and T. A. Witten
\textit{Contact line deposits in an evaporating drop},
Phys. Rev. E \textbf{83}, 026306 (2000).

\bibitem{Berteloot_epl08}
G. Berteloot, C.-T. Pham, A. Daerr, F. Lequeux and L. Limat,
\textit{Evaporation-induced flow near a contact line: Consequences on coating and contact angle}
EPL \textbf{83}, 14003 (2008).

\bibitem{Monteux_epl09}
C. Monteux, Y. Elmaallem, T. Narita and F. Lequeux,
\textit{Advancing-drying droplets of polymer solutions: Local increase of the viscosity at the contact line},
EPL \textbf{83} 34005 (2008).

\bibitem{Biance04}
A.-L. Biance, C. Clanet, and D. Qu\'er\'e,
\textit{First steps in the spreading of a liquid droplet,}
Phys. Rev. E \textbf{69}, 016301 (2004).

\bibitem{Bird08}
J. C. Bird, S. Mandre, and H. A. Stone,
\textit{Short-time dynamics of partial wetting,}
Phys. Rev. Lett. \textbf{100}, 234501 (2008).

\bibitem{Amberg12}
A. Carlson, G. Bellani, and G. Amberg,
\textit{Universality in dynamic wetting dominated by contact line friction,}
Phys. Rev. E \textbf{85}, 045302(R) (2012).

\bibitem{Winkels12}
K.G. Winkels, J.H. Weijs, A. Eddi, and J.H. Snoeijer,
\textit{Initial spreading of low-viscosity drops on partially wetting surfaces},
Phys. Rev. E \textbf{85}, 055301(R) (2012).

\bibitem{Eddi13}
A. Eddi, K.G. Winkels and J.H. Snoeijer,
\textit{Short time dynamics of viscous drop spreading,}
Phys. Fluids \textbf{25}, 013102 (2013).

\bibitem{Prudhomme96}
R.K. Prudhomme, G. Wu and D.K. Schneider,
\textit{Structure and rheology studies of Poly(oxyethylene-oxypropylene-oxyethylene) aqueous solution,}
Langmuir \textbf{12}, 4651-4659 (1996).

\bibitem{Habas04}
J.P. Habas, E. Pavie, A. Lapp and J. Peyrelasse,
\textit{Understanding the complex rheological behavior of PEO-PPO-PEO copolymers in aqueous solution,}
J. Rheol. \textbf{48}, 1-21 (2004).

\bibitem{Chaibundit07}
C. Chaibundit, N. M. P. S. Ricardo, F. de M. L. L. Costa, S. G. Yeates and C. Booth,
\textit{Micellization and Gelation of Mixed Copolymers P123 and F127 in Aqueous Solution,}
Langmuir \textbf{23}, 9229-9236 (2007).

\bibitem{Boucenna10}
I. Boucenna, L. Royon, P. Colinart, M.-A. Guedeau-Boudeville and A. Mourchid,
\textit{Structure and Thermorheology of Concentrated Pluronic Copolymer Micelles in the Presence of Laponite Particles,}
Langmuir \textbf{26}, 14430-14436 (2010).

\bibitem{Jalaal_Cottrell2017}
M. Jalaal, G. Cottrell, N.J. Balmforth and B. Stoeber,
\textit{On the rheology of Pluronic F127 aqueous solutions,}
J. Rheology. \textbf{61}, 139 (2017).

\bibitem{Tanner79}
L. Tanner, 
\textit{The spreading of silicon oil drops on horizontal surfaces,}
J. Phys. D: Appl. Phys. \textbf{12}, 1473 (1979).

\bibitem{Hocking94}
L. M. Hocking, 
\textit{The spreading of drops with intermolecular forces,} 
Phys. Fluids \textbf{6}, 3224-3228 (1994).

\bibitem{Stapelbroek14}
B.B.J. Stapelbroek, H.P. Jansen, E. S. Kooij, J. H. Snoeijer and A. Eddi,
\textit{Universal spreading of water drops on complex surfaces,}
Soft Matt. \textbf{10}, 2641-2648 (2014).

\bibitem{Davis_Solidification}
S. H. Davis,
\textit{Theory of solidification},
Cambridge University Press (2001).

\bibitem{Gelderblom_pre11}
H. Gelderblom, A. G. Marin, H. Nair, A. van Houselt, L. Lefferts, J. H. Snoeijer and D. Lohse,
\textit{How water droplets evaporate on a superhydrophobic substrate},
Phys. Rev. E \textbf{83}, 026306 (2011).

\bibitem{estimation}
The various quantities can  be extracted from experiments and from handbooks: $D$ = 24.6 $\times$ 10$^{-6}$ m$^2$.s$^{-1}$, $\rho_l$ = 1000 kg.m$^{-3}$, $c_s$ = 2.08 $\times$ 10$^{-2}$ kg.m$^{-3}$, while in the experiment RH = 0.5.  Hence, the velocity $U_0 = D c_s \textrm{RH}/R\rho_l \sim 2.5\times10^{-7}$ m.s$^{-1}$.


\end{thebibliography}
\end{document}